\documentstyle[preprint,aps,axodraw]{revtex}
\tightenlines

\newcommand{\nn}{\nonumber}
\newcommand{\ovl}[1]{\overline{#1}}
\newcommand{\pslash}{p\kern-1ex /}
\newcommand{\bpsi}{\overline{\psi}}
\def\setcaption#1{\def\@captype{#1}}

\begin{document}

%\draft

\title{
\vspace{-3.0cm}
\begin{flushright}  
{\normalsize UTCCP-P-33}\\
{\normalsize MPI-PhT/98-15}\\
{\normalsize UTHEP-380}\\
\end{flushright}
Perturbative Renormalization Factors of Bilinear Quark Operators 
for Improved Gluon and Quark Actions in Lattice QCD}

\author{$^{1,2}$Sinya Aoki, $^3$Kei-ichi Nagai, $^1$Yusuke Taniguchi 
and $^1$Akira Ukawa} 

\address{$^1$Institute of Physics, University of Tsukuba, 
Tsukuba, Ibaraki-305, Japan \\
$^2$Max-Planck-Institut f\"ur Physik, F\"oringer Ring 6, D-80805 
M\"unchen, Germany\\
$^3$Center for Computational Physics, University of Tsukuba,
Tsukuba, Ibaraki-305, Japan \\}

\date{\today}

\maketitle

\begin{abstract}
We calculate one-loop renormalization factors of bilinear quark 
operators for gluon action including six-link loops and $O(a)$-improved
quark action in the limit of massless quark.  
We find that finite parts of one-loop coefficients of 
renormalization factors diminish monotonically as either of 
the coefficients $c_1$ or $c_2+c_3$ 
of the six-link terms are decreased below zero.  
Detailed numerical results are 
given, for general values of the clover coefficient, 
for the tree-level improved gluon action in the Symanzik 
approach $(c_1=-1/12, c_2=c_3=0)$ and for the choices suggested 
by Wilson $(c_1=-0.252, c_2=0, c_3=-0.17)$ and by Iwasaki 
$(c_1=-0.331, c_2=c_3=0$ and $c_1=-0.27, c_2+c_3=-0.04)$ from 
renormalization-group analyses. 
Compared with the case of the standard plaquette gluon action, 
finite parts of one-loop coefficients are reduced by 10--20\% 
for the Symanzik action, 
and approximately by a factor two for the renormalization-group 
improved gluon actions.   

\end{abstract}


\narrowtext

\section{Introduction}

Lattice QCD calculations of hadron matrix elements require 
values of renormalization factors which relate operators 
on the lattice to those in the continuum.  For the standard 
plaquette gluon action, perturbative calculation of renormalization 
factors for massless quark has been carried out to one-loop order 
for bilinear and four-quark operators,  both for the Wilson quark 
action\cite{Arroyo82,Martinelli83,Groot84,Martinelli84,Bernard87,Curci88} and 
for the $O(a)$-improved ``clover'' action
\cite{Hamber83,Gabrielli91,Frezzotti92,Borrelli93,Luescher96,Sint97}
originally suggested by Sheikholeslami and Wohlert\cite{SW85,Heatlie91}.
With developments of our full QCD simulations\cite{CPPACS} 
employing an improved gluon action\cite{Iwasaki83}, 
we find it necessary to extend these calculations.  
In this article we report results for renormalization 
factors of bilinear quark operators at the one-loop level for the gluon 
action improved by addition of six link loops to the plaquette term.   
For the quark action we treat both the Wilson and $O(a)$-improved actions, 
taking the limit of massless quark.  
We evaluate numerical values of the one-loop coefficients of renormalization 
factors for the case of tree-level improved action in the Symanzik 
approach\cite{Weisz83,Luescher85}, and for several cases of 
renormalization-group improved actions\cite{Iwasaki83,Wilson80}.  
We also examine how the one-loop coefficients vary for general values 
of the coefficients of the six-link loop terms.

In Sec.~II we write down the action we treat and the Feynman rule 
to fix our conventions.  
The structure of renormalization factors related to fermion self-energy is   
discussed in Sec.~III, and that for bilinear quark operators in Sec.~VI.
Numerical results for one-loop coefficients are given in Sec.~V. 
Since the procedure of calculation is by now standard, we shall be brief 
on this point.  Expressions for one-loop integrands are listed in 
Appendix A. In Appendix B we collect one-loop results for the relation 
between the renormalized and bare coupling constants for various choices 
of gluon and quark actions.  

\section{Action and Feynman rule}

The gluon action we consider is defined by 
\begin{equation}
S_{\rm gluon} = \frac{1}{g^2}\left\{
c_0 \sum_{plaquette} {\rm Tr}U_{pl}
+ c_1  \sum_{rectangle} {\rm Tr} U_{rtg}
+ c_2 \sum_{chair} {\rm Tr} U_{chr}
+ c_3 \sum_{parallelogram} {\rm Tr} U_{plg}\right\}, 
\end{equation}
where the first term represents the standard plaquette term, and the 
remaining terms are six-link loops formed by a $1\times 2$ rectangle, 
a bent $1\times 2$ rectangle (chair) and a 3-dimensional parallelogram. 
The coefficients $c_0, \cdots, c_3$ satisfy the normalization condition
\begin{equation}
c_0+8c_1+16c_2+8c_3=1. 
\end{equation}

For the quark action we consider
\begin{equation}
S_{\rm quark} = S_{\rm W} + S_{\rm C}, 
\end{equation}
where $S_{\rm W}$ is the Wilson action given by 
\begin{eqnarray}
S_{\rm W}=
a^3 \sum_n \frac{1}{2} \sum_\mu \left(
\bpsi_n (-r + \gamma_\mu) U_{n, \mu} \psi_{n+\hat{\mu}}
+
\bpsi_n (-r - \gamma_\mu) U^\dagger_{n-\mu, \mu}
\psi_{n-\hat{\mu}} \right)
+ (a m_0+4r) \bpsi_n \psi_n  
\end{eqnarray}
with $a$ the lattice spacing, 
and $S_{\rm C}$ represents the ``clover'' term defined by 
\begin{eqnarray}
S_{\rm C} = - c_{\rm SW} a^3 \sum_n \sum_{\mu, \nu}
ig \frac{r}{4} \bpsi_n \sigma_{\mu \nu} P_{\mu \nu} (n) \psi_n . 
\end{eqnarray}
For the field strength $P_{\mu\nu}$ we adopt the definition given by 
\begin{eqnarray}
&&
P_{\mu \nu} (n) = \frac{1}{4} \sum_{i=1}^4 \frac{1}{2ig}
\left( U_i (n) - U^\dagger_i (n) \right) , 
\\
&&
U_1 (n) = U_{n, \mu} U_{n+\hat{\mu}, \nu}
 U^\dagger_{n+\hat{\nu}, \mu} U^\dagger_{n, \nu} , 
\\
&&
U_2 (n) = U_{n, \nu} U^\dagger_{n-\hat{\mu}+\hat{\nu}, \mu}
 U^\dagger_{n-\hat{\mu}, \nu} U_{n-\hat{\mu}, \mu} , 
\\
&&
U_3 (n) = U^\dagger_{n-\hat{\mu}, \mu}
 U^\dagger_{n-\hat{\mu}-\hat{\nu}, \nu} U_{n-\hat{\mu}-\hat{\nu}, \mu}
 U_{n-\hat{\nu}, \nu} , 
\\
&&
U_4 (n) = U^\dagger_{n-\hat{\nu}, \nu}
 U_{n-\hat{\nu}, \mu} U_{n+\hat{\mu}-\hat{\nu}, \nu}
 U^\dagger_{n, \mu} .  
\end{eqnarray}
Our $\gamma$ matrix convention is as follows:
\begin{eqnarray}
&&\gamma_i
= \pmatrix{
  0        & -i \sigma^i \cr
i \sigma^i & 0           \cr
},
\quad
\gamma_4
= \pmatrix{
0 & 1 \cr
1 & 0 \cr
} , \\
&&\gamma_5 \equiv \gamma_1 \gamma_2 \gamma_3 \gamma_4
=\pmatrix{
1 &  0 \cr
0 & -1 \cr
} , \\
&&
\sigma_{\mu \nu} = \frac{1}{2} \left[ \gamma_\mu , \gamma_\nu \right] . 
\end{eqnarray}

Weak-coupling perturbation theory is developed by writing
\begin{equation}
U_{n,\mu}=\exp (igaA_\mu(n+\frac{1}{2}\hat\mu)) .
\end{equation}
We adopt a covariant gauge fixing with a gauge parameter $\alpha$ 
defined by  
\begin{eqnarray}
S_{\rm GF} = a^4 \sum_n \frac{1}{2\alpha}
\left( \nabla_\mu A_\mu^a (n+\frac{1}{2}\hat{\mu}) \right)^2 ,  
\end{eqnarray}
where $\nabla_\mu f_n\equiv (f_{n+\hat{\mu}}-f_n)/a$. 
The free part of the gluon action takes the form
\begin{equation}
S_0 = \frac{1}{2} \int_{-\pi/a}^{+\pi/a}\frac{d^4k}{(2\pi)^4}
\sum_{\mu, \nu}
A_\mu^a(k) G_{\mu \nu} A_\nu^a(-k) , 
\end{equation}
where
\begin{equation}
G_{\mu \nu} = \hat{k}_\mu \hat{k}_\nu + \sum_\rho 
(\hat{k}_\rho \delta_{\mu \nu} - \hat{k}_\mu \delta_{\rho
\nu}) q_{\mu \rho} \hat{k}_\rho
\end{equation}
with 
\begin{equation}
\hat{k}_\mu = \frac{2}{a} \sin \frac{k_\mu a}{2}
\end{equation}
and $q_{\mu \nu}$ is defined as
\begin{equation}
q_{\mu \nu} = (1-\delta_{\mu\nu})\left(1 - (c_1 - c_2 - c_3)a^2
(\hat{k}_\mu^2 + \hat{k}_\nu^2) -(c_2 + c_3)a^2 \hat{k}^2\right) .
\end{equation}
The gluon propagator can be written as 
\begin{eqnarray}
D_{\mu \nu}& =& (\hat{k}^2)^{-2} \left[
\hat{k}_\mu \hat{k}_\nu  + \sum_\sigma 
(\hat{k}_\sigma \delta_{\mu \nu} - \hat{k}_\nu \delta_{\mu
\sigma} ) \hat{k}_\sigma A_{\sigma \nu} \right]
-(1-\alpha) \frac{\hat{k}_\mu \hat{k}_\nu}{(\hat{k}^2)^2}
\\
& = & (\hat{k}^2)^{-2} \left[
(1 - A_{\mu \nu} )\hat{k}_\mu \hat{k}_\nu 
+ \delta_{\mu \nu} \sum_\sigma \hat{k}_\sigma^2
A_{\nu \sigma} \right]
-(1-\alpha) \frac{\hat{k}_\mu \hat{k}_\nu}{(\hat{k}^2)^2} , 
\end{eqnarray}
where $A_{\mu \nu}$ is a function of $q_{\mu \nu}$ and
$\hat{k}_\mu$ whose form we refer to the original 
literatures\cite{Iwasaki83,Weisz83}. 

The free quark propagator is given by 
\begin{eqnarray}
S_{\rm q}(p) =
\frac{ -i \sum_\mu \gamma_\mu \ovl{p}_\mu + W(p)}
{\sum_\mu\ovl{p}_\mu^2 + W(p)^2} , 
\end{eqnarray}
where 
\begin{eqnarray}
\bar{k}_\mu &=& \frac{1}{a} \sin k_\mu a\\
W(p) &=& m+\frac{r}{a}\sum_\mu (1-\cos a p_\mu).
\end{eqnarray}

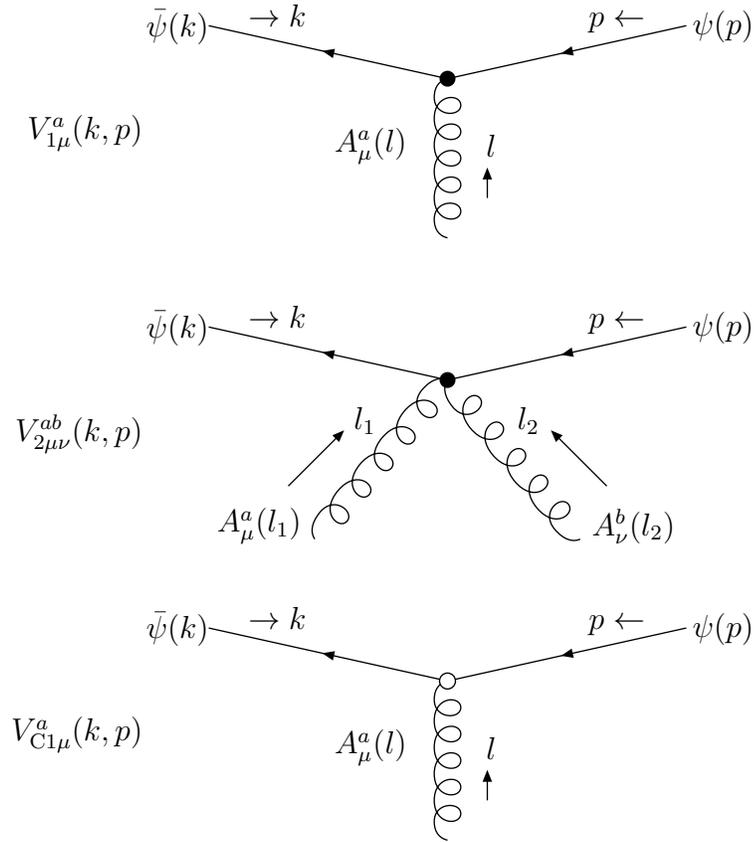
\begin{figure}
\begin{center}\begin{picture}(200,100)(0,0)
\Text(-15,50)[r]{$V_{1\mu}^a(k,p)$}
\ArrowLine(100,70)(10,90)
\Text(10,90)[r]{$\bar\psi(k)$}
\Text(25,90)[lb]{$\rightarrow k$}
\ArrowLine(190,90)(100,70)
\Text(193,90)[l]{$\psi(p)$}
\Text(175,90)[rb]{$p \leftarrow$}
\Gluon(100,10)(100,70){5}{5}
\Text(115,45)[l]{$l$}\LongArrow(115,25)(115,35)
\Text(85,45)[r]{$A^a_\mu(l)$}
\Vertex(100,70){3}
\end{picture}\end{center}
\begin{center}\begin{picture}(200,100)(0,0)
\Text(-15,50)[r]{$V_{2\mu\nu}^{ab}(k,p)$}
\ArrowLine(100,70)(10,90)
\Text(10,90)[r]{$\bar\psi(k)$}
\Text(25,90)[lb]{$\rightarrow k$}
\ArrowLine(190,90)(100,70)
\Text(193,90)[l]{$\psi(p)$}
\Text(175,90)[rb]{$p \leftarrow$}
\Gluon(50,10)(100,70){5}{5}
\Text(65,55)[l]{$l_1$}\LongArrow(40,30)(60,50)
\Text(45,15)[r]{$A^{a}_\mu(l_1)$}
\Gluon(150,10)(100,70){5}{5}
\Text(135,55)[r]{$l_2$}\LongArrow(160,30)(140,50)
\Text(155,15)[l]{$A^{b}_\nu(l_2)$}
\Vertex(100,70){3}
\end{picture}\end{center}
\begin{center}\begin{picture}(200,100)(0,0)
\Text(-15,50)[r]{$V_{{\rm C}1\mu}^a(k,p)$}
\ArrowLine(100,70)(10,90)
\Text(10,90)[r]{$\bar\psi(k)$}
\Text(25,90)[lb]{$\rightarrow k$}
\ArrowLine(190,90)(100,70)
\Text(193,90)[l]{$\psi(p)$}
\Text(175,90)[rb]{$p \leftarrow$}
\Gluon(100,10)(100,70){5}{5}
\Text(115,45)[l]{$l$}\LongArrow(115,25)(115,35)
\Text(85,45)[r]{$A^a_\mu(l)$}
\BCirc(100,70){3}
\end{picture}
\caption[]{Quark-gluon vertices needed for our one-loop calculations.}
\label{fig:rule}
\end{center}
\end{figure}

To calculate renormalization factors of bilinear quark operators to 
one-loop order,  we need only one- and two-gluon vertices with quarks. 
The vertices originating from the Wilson quark action are given by 
\begin{eqnarray}
V_{1\mu}^a (k,p)
&=& -i g T^a \{ \gamma_\mu \cos \frac{a}{2}(-k_\mu + p_\mu)
  -i r \sin \frac{a}{2}(-k_\mu + p_\mu) \} , \\
V_{2\mu\nu}^{ab} (k,p)
&=& \frac{a}{2} g^2 \frac{1}{2} \{T^{a}, T^{b}\}
\{ i \gamma_\mu \sin \frac{a}{2}(-k_\mu + p_\mu)
-r \cos \frac{a}{2} (-k_\mu + p_\mu) \}\delta_{\mu\nu} , 
\end{eqnarray}
and the interaction due to the clover term has the form
\begin{eqnarray}
V_{{\rm C}1\mu}^a(k,p)=
-c_{\rm SW} g \frac{r}{2} \left( \sum_\nu \sigma_{\mu \nu}
\sin a(p_\nu + k_\nu) \right) \cos \frac{a}{2} (p_\mu + k_\mu) T^a . 
\end{eqnarray}
Our momentum assignments for the vertices are depicted in Fig.~\ref{fig:rule}.
The two-gluon vertex with quarks from the clover term gives no
contribution to diagrams we evaluate, and is hence omitted.

\section{Quark serf-energy}

Let us write the inverse full quark propagator as 
\begin{equation}
S_{\rm q}^{-1}(p)=i\gamma_\mu\bar p_\mu+W(p)-\Sigma (p) . 
\end{equation}
Calculating the quark self-energy $\Sigma (p)$ to one-loop order 
on the lattice and making an expansion of form 
\begin{equation}
\Sigma (p)=\Sigma (0) +\frac{\partial\Sigma (0)}{\partial p_\mu}p_\mu+O(p^2), 
\end{equation}  
we find 
\begin{eqnarray}
S_q^{-1}( p ) & = &
i \gamma_\mu p_\mu \left[
1-\frac{g^2}{16\pi^2} C_F\left\{ \log (\lambda a)^2 +\Sigma_1+
(\alpha-1)(\log (\lambda a)^2+\Sigma_\alpha)\right\}
\right]\nn \\
&&+ m
\left[ 1-\frac{g^2}{16\pi^2} C_F\left\{ 4 \log (\lambda a)^2 + \Sigma_2+
(\alpha-1)(\log (\lambda a)^2+\Sigma_\alpha)
\right\}\right], 
\label{eq:selfenergy}
\end{eqnarray}
where
\begin{equation}
m =  m_0 -\frac{g^2}{16\pi^2} C_F \frac{\Sigma_0}{a} .  
\end{equation}
In (\ref{eq:selfenergy}) the coefficients of $i\gamma_\mu p_\mu$ and $m$ 
are evaluated with $m_0=0$.  The infrared divergence that
appears in this case is regularized by a gluon mass $\lambda$ 
introduced in the gluon propagator as described in Appendix A. 
Of the finite constants $\Sigma_i (i=0, 1, 2, \alpha)$, 
$\Sigma_\alpha$ is independent of the clover coefficient $c_{SW}$ nor does 
it depend on the improved part of the gluon action.
The other constants $\Sigma_i (i=0, 1, 2)$ depend quadratically on the 
clover coefficient $c_{SW}$, and we write 
\begin{eqnarray}
\Sigma_i & = & \Sigma_i^{(0)}+ c_{SW}\Sigma_i^{(1)}+
c_{SW}^2 \Sigma_i^{(2)} .
\end{eqnarray}

Let us recall that the hopping parameter $K$ is related to the bare 
quark mass $m_0$ through 
\begin{equation}
K=\frac{1}{2(m_0a+4)}
\end{equation}
The critical hopping parameter corresponding to massless quark is given by
\begin{equation}
K_c = \frac{1}{8}\left(1-\frac{g^2}{16\pi^2}C_F \frac{\Sigma_0}{4}\right) .
\end{equation}

In the continuum we employ the $\overline{MS}$ scheme with naive dimensional 
regularization.  The one-loop self-energy in the continuum has the same form 
as (\ref{eq:selfenergy}) with, however, the replacements,  
\begin{eqnarray}
\log (\lambda a)^2&\to&\log (\lambda/\mu)^2 , \\
\Sigma_1 & \to & \Sigma_1^{\overline{MS}}=1/2 , \\
\Sigma_2 & \to & \Sigma_2^{\overline{MS}}=-2 , \\
\Sigma_\alpha & \to & \Sigma_\alpha^{\overline{MS}}=-1 . 
\end{eqnarray}

Let us define the quark wave function renormalization factor needed for 
converting the lattice field to the continuum field in the $\overline{MS}$ 
scheme by 
\begin{equation} 
\psi^{\overline{MS}}= Z_\psi \psi^{lat} . 
\end{equation}
To one-loop order we then find that
\begin{equation}
Z_\psi (\mu a) = 1+\frac{g^2}{16\pi^2}C_F\left[ 
-\log (\mu a)^2 +z_\psi 
+(\alpha -1)(-\log (\mu a)^2)\right] .
\end{equation}
where 
\begin{equation}
z_\psi = \Sigma_1^{\overline{MS}} -\Sigma_1
+(\alpha -1)(\Sigma_\alpha^{\overline{MS}} -\Sigma_\alpha) .
\end{equation}
The quark mass renormalization factor defined by
\begin{equation}
 m^{\overline{MS}}(\mu ) = Z_m (\mu a) m
\end{equation}
is given by
\begin{equation}
Z_m (\mu a) = 1+\frac{g^2}{16\pi^2}C_F\left[ 
-3 \log (\mu a)^2 +z_m \right] .
\end{equation}
with 
\begin{equation}
z_m = (\Sigma_2^{\overline{MS}} -\Sigma_2)
    - (\Sigma_1^{\overline{MS}} -\Sigma_1) .
\end{equation}
It is worth noting that the critical hopping parameter $K_c$ and 
the quark mass renormalization factor $Z_m$ do not depend on 
the gauge parameter $\alpha$.

\section{Bilinear operators}

We consider bilinear quark operators in the following form\cite{Borrelli93}:
\begin{equation}
O_\Gamma (z) = [1 + r am_0 (1-z)]\bar\psi\Gamma\psi + 
z\bar\psi\Gamma^\otimes\psi
-z^2 \bar\psi\Gamma^\prime\psi , 
\label{eq:operator}
\end{equation}
Here $\Gamma$ denotes appropriate $\gamma$ matrices, and the additional 
operators in the second and the third term 
needed for improvement are defined by 
\begin{eqnarray}
\bar\psi\Gamma^\otimes\psi &\equiv& \frac{r}{4}
\left[ (\bar\psi_{n+\mu}U_{n,\mu}^\dagger-\bar\psi_{n-\mu}U_{n-\mu,\mu})
\gamma_\mu \Gamma\psi_n  
-\bar\psi_n\Gamma\gamma_\mu(U_{n,\mu}\psi_{n+\mu}-U_{n-\mu,\mu}^\dagger
\psi_{n-\mu})\right] , \\
\bar\psi\Gamma^\prime\psi &\equiv&
\frac{r^2}{16}
\left[ (\bar\psi_{n+\mu}U_{n,\mu}^\dagger-\bar\psi_{n-\mu}U_{n-\mu,\mu})
\gamma_\mu \Gamma\gamma_\nu(U_{n,\nu}\psi_{n+\nu}-U_{n-\nu,\nu}^\dagger
\psi_{n-\nu})\right] .
\end{eqnarray}

At the one-loop order, on-shell matrix elements of $O_\Gamma (z)$ do not 
have terms of $O(a)$ or $O(g^2a\log a)$\cite{Heatlie91}.  To remove 
terms of $O(g^2a)$,  the coefficients of the added operators 
as well as that of $am_0$ in (\ref{eq:operator}) have to be corrected 
by an $O(g^2)$ term. For vector and axial vector currents, an addition of 
a total derivative operator with an $O(g^2)$ coefficient is also 
necessary\cite{Luescher96}.  
Evaluating these coefficients is outside the scope of this article. 

We calculate the renormalization factor of $O_\Gamma (z)$ for massless quark. 
The on-shell vertex functions for the three operators in (\ref{eq:operator}), 
for a quark and an antiquark with momenta $p=p^\prime=0$ 
as external states, take the form, 
\begin{eqnarray}
T_\Gamma &=& \left[ 1 +\frac{g^2 C_F}{16\pi^2} \left\{ 
-\frac{h_2(\Gamma)}{4}\log(\lambda a)^2+V_\Gamma  
+(\alpha - 1) \left(-\log(\lambda a)^2+V_\alpha \right)\right\}
\right]\Gamma\\
T_{\Gamma^\otimes} &=& \frac{g^2 C_F}{16\pi^2} V_{\Gamma^\otimes} \Gamma ,\\
T_{\Gamma^\prime} &=& \frac{g^2 C_F}{16\pi^2} V_{\Gamma^\prime} \Gamma ,
\end{eqnarray}
where $\lambda$ is the gluon mass, and $h_2(\Gamma)$ is an integer given by 
\begin{equation}
h_2(\Gamma)=4(A), 4(V), 16(P), 16(S), 0(T)
\end{equation}
for various Dirac channels. The finite constants 
$V_i (i=\Gamma, \Gamma^\otimes, \Gamma^\prime)$ depend quadratically on
the clover coefficients $c_{SW}$, and we write
\begin{eqnarray}
V_\Gamma &=& V_\Gamma^{(0)}+ c_{SW}\cdot V_\Gamma^{(1)} 
+ c_{SW}^2 V_\Gamma^{(2)} , \\
V_{\Gamma^\otimes} &=& V_{\Gamma^\otimes}^{(0)}
+ c_{SW}\cdot V_{\Gamma^\otimes}^{(1)} + c_{SW}^2 V_{\Gamma^\otimes}^{(2)} ,\\
V_{\Gamma^\prime} &=& V_{\Gamma^\prime}^{(0)}
+ c_{SW}\cdot V_{\Gamma^\prime}^{(1)} + c_{SW}^2 V_{\Gamma^\prime}^{(2)} .
\end{eqnarray}
The constant $V_\alpha$ satisfies $V_\alpha=-\Sigma_\alpha$.

In the continuum, the on-shell vertex function to one-loop order 
is given in the $\overline{MS}$ scheme by 
\begin{equation}
T_\Gamma^{\overline{MS}} = \left[ 1 +\frac{g^2 C_F}{16\pi^2} \left\{
\frac{h_2(\Gamma)}{4} \log(\mu/\lambda)^2 +V_\Gamma^{\overline{MS}} +
(\alpha - 1) \left(\log(\mu/\lambda)^2 + 1\right)\right\}
\right]\Gamma
\end{equation}
with
\begin{equation}
V_\Gamma^{\overline{MS}} =  \frac{3 h_2(\Gamma)}{8}+j(\Gamma) ,  
\end{equation}
where 
\begin{equation}
j(\Gamma)=-2(-6)(A), -2(-2)(V), -4(4)(P), -4(-4)(S), 0(0)(T)
\end{equation}
for the anti-commuting and 'tHooft-Veltman definition of $\gamma_5$, with 
values for the latter in parentheses.

Combining the above results and including self-energy corrections, 
the relation between the continuum operator in the $\overline{MS}$ scheme
and the lattice operator is given by 
\begin{equation}
O_\Gamma^{\overline{MS}}(\mu)=Z_\Gamma\left[
        \left[1+rma(1-z)\right]\bar\psi\Gamma\psi
        +z\bar\psi\Gamma^\otimes\psi
        -z^2\bar\psi\Gamma^\prime\psi\right] ,
\label{eq:renop}
\end{equation}
where 
\begin{equation}
Z_\Gamma = 1+\frac{g^2 C_F}{16\pi^2}\left[ \left(\frac{h_2(\Gamma)}{4}
-1\right)\log(\mu a)^2 + z_\Gamma \right]
\end{equation}
and 
\begin{equation}
z_\Gamma = Z_\Gamma^{\overline{MS}}-Z_\Gamma^{lat}
\end{equation}
with  
\begin{eqnarray}
Z_\Gamma^{\overline{MS}} &=& \frac{3 h_2(\Gamma)}{8}+j(\Gamma)+1/2 , \\
Z_\Gamma^{lat} &=& V_\Gamma + \Sigma_1+(1-z)r\Sigma_0
                +z V_{\Gamma^\otimes} -z^2 V_{\Gamma^\prime} .
\end{eqnarray}

For evaluating values of renormalization factors, estimating 
the renormalized coupling constant for a given value of the bare 
coupling constant often becomes necessary.  We collect one-loop results 
needed for such an estimation for various choices of gluon and quark 
actions in Appendix B. 

\section{Numerical results}

\subsection{Calculational procedure}

We calculate renormalization factors of bilinear quark operators 
for massless quark with the Wilson parameter $r=1$.  
Two methods are employed to calculate the finite parts of lattice 
amplitudes.  In the first method Dirac algebra of Feynman integrands 
are carried out by hand, and the momentum integration is performed 
by a mode sum for a periodic box of a size $L^4$ after transforming 
the momentum variable through $p_\mu=q_\mu-\sin q_\mu$.  
We employ the size $L=48$ for integrals which are infra-red finite, and sizes 
up to $L=128$ for those whose infra-red divergence is regularized by 
subtraction of the leading singular terms for small loop momenta. 
In the second method, a {\it Mathematica} 
program is written to perform Dirac algebra.  
The output is converted into a FORTRAN code, also by 
{\it Mathematica}, and the momentum integration is carried out by the 
Monte Carlo routine VEGAS, using 10 samples of 50000 points each.  
The agreement of results from the two methods 
is used as a check of our calculations. 
In Appendix A we list explicit forms of integrands, and explain how 
we regularize infrared divergence. 

\subsection{Results for representative gluon actions}

At the one-loop level, the choice of the gluon action is specified by the 
pair of numbers $c_1$ and $c_{23}=c_2+c_3$.  As representative cases, 
we report numerical results for renormalization factors for 
(i) the tree-level improved action in the Symanzik approach 
$c_1=-1/12, c_{23}=0$\cite{Weisz83,Luescher85}, 
and (ii) three choices suggested 
by an approximate renormalization-group analysis, $c_1=-0.331, c_{23}=0$ 
and $c_1=-0.27, c_{23}=-0.04$ by Iwasaki\cite{Iwasaki83}, 
and $c_1=-0.252, c_{23}=-0.17$ by Wilson\cite{Wilson80}. 
We also include results for the standard plaquette action 
$c_1=0, c_{23}=0$ to facilitate comparison with the cases above, 
and also as a check of our results with those in literature
\cite{Arroyo82,Martinelli83,Groot84,Gabrielli91,Frezzotti92,Borrelli93,Sint98}.

Since one-loop renormalization factors are quadratic polynomials in 
the clover coefficient $c_{SW}$, we tabulate results 
for the coefficients of the polynomial. 

In Table~\ref{tab:self} we list the finite constants for quark self-energy
for the choices of gluon action described above.  The contribution of 
the tadpole diagram is also listed for $\Sigma_0$ and $\Sigma_1$. 
Errors from numerical integration are at most in the last digit written. 
Combining values in this table, we find the finite part $z_m$ 
for the quark mass renormalization factor, which is  
tabulated in Table~\ref{tab:qmass}. 

Results for finite parts for lattice bilinear operators are given in 
Table~\ref{tab:local} ($V_\Gamma$), \ref{tab:impr} ($V_{\Gamma^\otimes}$)
and \ref{tab:impr_a2} ($V_{\Gamma^\prime}$).  
For local operators ($z=0$ in (\ref{eq:operator}) and (\ref{eq:renop})), 
these values lead to 
the finite part $z_\Gamma$ listed in Table~\ref{tab:zfact} where we 
adopt the anti-commuting definition for $\gamma_5$. 

Looking at numerical values in these tables we observe that 
the one-loop coefficients are reduced
by 10--20\% for the tree-level Symanzik action compared to those for the 
plaquette action.  The reduction of magnitude is enhanced to a factor 
of about two for the renormalization-group improved actions.

\subsection{Dependence on $c_1$ and $c_2+c_3$}

\begin{figure}
\epsfxsize=6cm
\begin{center}
\setcaption{figure}
\ \epsfbox{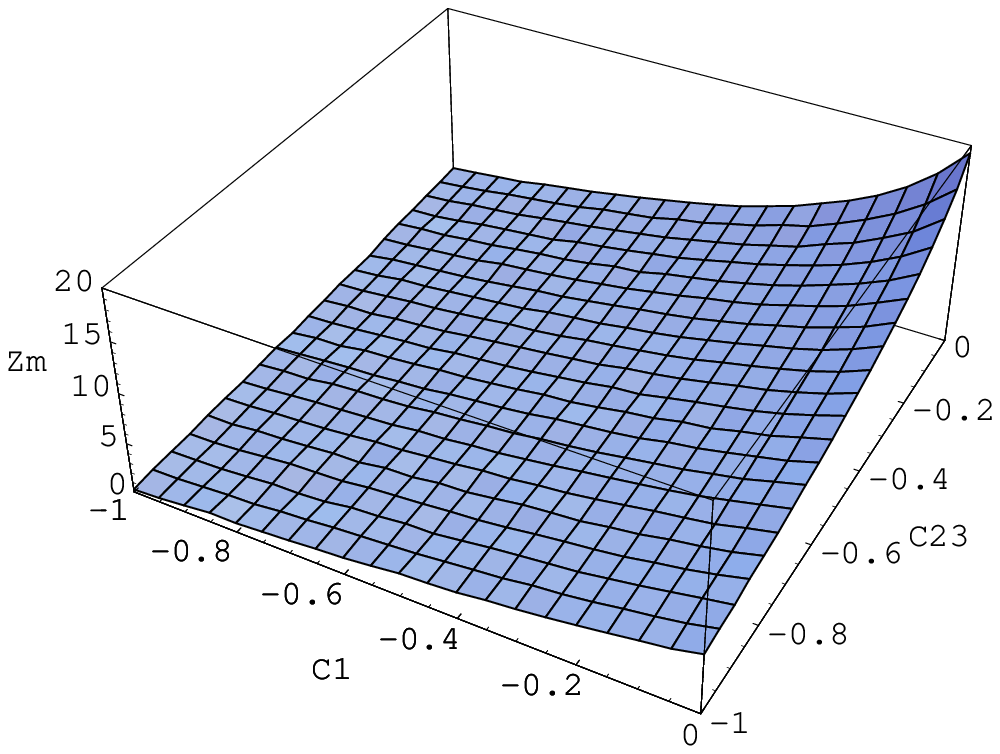}
\caption[]{Lattice finite part $z_m$ for quark mass renormalization factor 
as a function of $c_1$ and $c_{23}$ for clover quark action with
$c_{SW}=1$.}
\label{fig:zm}
\end{center}

\epsfxsize=6cm
\begin{center}
\setcaption{figure}
\ \epsfbox{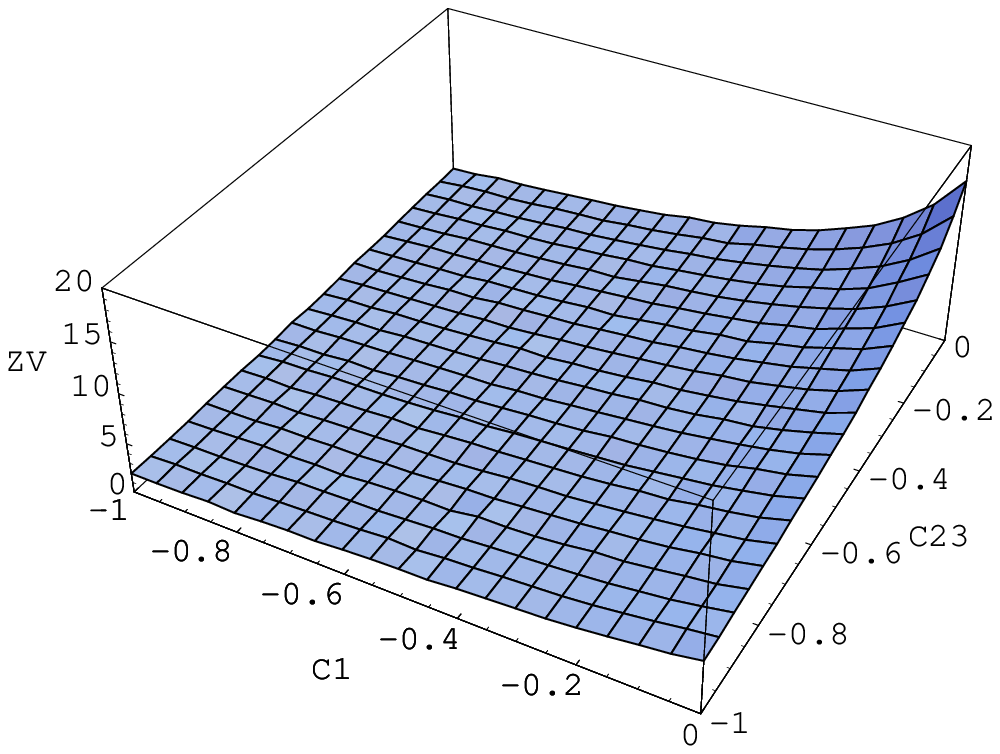}
\caption[]{Lattice finite part $z_V$ for vector current 
as a function of $c_1$ and $c_{23}$ for clover quark action with
$c_{SW}=1$. }
\end{center}

\epsfxsize=6cm
\begin{center}
\setcaption{figure}
\ \epsfbox{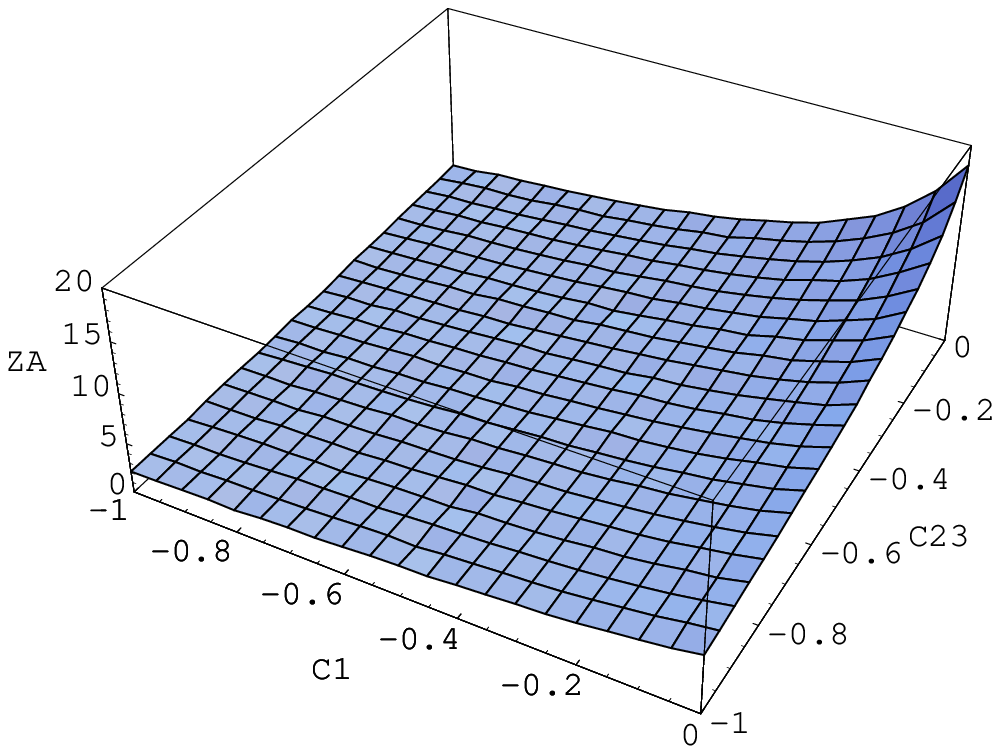}
\caption[]{Lattice finite part $z_A$ for axial vector current
as a function of $c_1$ and $c_{23}$ for clover quark action with
$c_{SW}=1$.}
\label{fig:za}
\end{center}

\end{figure}

The results above lead to a natural question how renormalization factors 
vary on the $(c_1, c_{23})$ plane.   To examine this point, we calculate 
the lattice part of one-loop coefficients for $-1\leq c_1, c_{23}\leq 0$
in steps of 0.05.  Choosing the tree level value $c_{SW}=1$ and 
adding the three terms of the quadratic polynomial in $c_{SW}$, 
we find the results plotted in Figs.~\ref{fig:zm} -- \ref{fig:za}. 
These plots show that the renormalization factors are monotonic
functions of $c_1$ and $c_{23}$, becoming smaller as either $c_1$ or 
$c_{23}$ is decreased below zero.  

\section*{Acknowledgements}

Part of numerical calculations for the present work has been carried 
out at Center for Computational Physics, University of Tsukuba. 
S. A. thanks Dr. S. Sint for providing us with his unpublished report.
This work is supported in part by the Grants-in-Aid for
Scientific Research from the Ministry of Education, Science and Culture
(Nos. 2373, 08640349, 08640350, 09246206).  
Y. T. is supported by Japan Society for Promotion of Science.
 
\section*{Appendix A}

In this appendix we list explicit forms of one-loop integrands 
for the quark self-energy and vertex functions.
The lattice spacing is set $a=1$ for simplicity. 

We first consider the quark self-energy.
Using the same notation as in text we find 
\begin{eqnarray}
\Sigma_0^{(n)} &=& -16\pi^2\left[ 4r T\delta_{n,0} 
+\int \frac{1}{F_0}I_0^{(n)}\frac{1}{G_0} \right]
\nn
\end{eqnarray}
for $n=$0, 1, 2, 
where
\begin{eqnarray}
\int &=& \int_{-\pi}^\pi \frac{d^4 k}{(2\pi)^4},\qquad
G_0  = (\hat{k}^2)^2 \qquad
F_0  =  \sum_\nu \sin^2k_\nu+ M^2,\qquad
W = r\sum_\nu (1-\cos k_\nu )  . \nn
\end{eqnarray}
Here $T$ represents the tadpole integral for the improved action, 
which is given by
\begin{eqnarray}
T &=& \int \frac{\sum_\nu \hat{k}_\nu^2 \bar{A}_{\mu\nu}}{2G_0} ,\nn
\end{eqnarray}
where $\bar{A}_{\mu\nu} =\delta_{\mu\nu}+A_{\mu\nu}$ with $A_{\mu\nu}$ 
as defined in text.
The integrand $I_0^{(n)}$ is expressed as
\begin{eqnarray}
I_0^{(0)}&=&  -8r^3 \Delta_1^3 -16r\Delta_3\Delta_1+ 32r\Delta_1 
\Delta_{1,0}^\mu , 
\nn \\
I_0^{(1)}&=& 16\Delta_3^2 -4\Delta_{1,1}^\mu
-16(4\Delta_3-s_\mu^2)\Delta_{1,0}^\mu , 
\nn \\
I_0^{(2)}&=& \frac{r}{2}\Delta_1I_0^{(1)} , 
\nn
\end{eqnarray}
where we define
\begin{eqnarray}
s_\mu &=&\sin k_\mu,\quad
c_\mu =\cos k_\mu,\quad
c=\sum_\nu c_\nu,\quad
\Delta_{1}= \frac{1}{4}\hat{k}^2,\quad 
\Delta_{3}=\frac{1}{4}\sum_\nu s_\nu^2,\quad
\Delta_{4}=\sum_\nu c_\nu s_\nu^2, 
\nn \\
\Delta_{1,1}^\mu &=&\sum_\nu\bar{A}_{\mu\nu}s_\mu^2 s_\nu^2
\quad
\Delta_{1,0}^\mu =\sum_\nu\bar{A}_{\mu\nu}\cos^2\frac{k_\mu}{2} 
\sin^2\frac{k_\nu}{2} .
\nn
\end{eqnarray}
Note that no sum is taken over the index $\mu$ for $\Delta_{1,1}^\mu$ 
and $\Delta_{1,0}^\mu$. 

Similarly we find 
\begin{eqnarray}
\Sigma_2^{(n)}&=&-16\pi^2\int \left[
\frac{1}{F_0}\left\{I_2^{(n)}-\frac{4r\Delta_1 I_0^{(n)}}{F_0}\right\}
\frac{1}{G_0}
-\frac{4}{(k^2)^2}\theta(\pi^2-k^2)\delta_{n,0}
\right]
-4\log \pi^2\delta_{n,0} , \nn
\end{eqnarray}
where
\begin{eqnarray}
I_2^{(0)}&=&-4r^2\Delta_1^2 +16\Delta_{1,0}^\mu,\quad
I_2^{(1)} = 0,\quad
I_2^{(2)} = \frac{I_0^{(1)}}{4},
\nn
\end{eqnarray}
and 
\begin{eqnarray}
\Sigma_\alpha&=&-16\pi^2\int \left[
\frac{1}{G_0}
-\frac{1}{(k^2)^2}\theta(\pi^2-k^2)
\right]
-\log \pi^2 . \nn
\end{eqnarray}
These formula illustrate how we regularize infrared divergence. 

Finally we have 
\begin{eqnarray}
\Sigma_1^{(n)} &=&16\pi^2 T\delta_{n,0} -
16\pi^2\int\left[ \frac{1}{F_0}\left\{ I_1^{(n)}+\frac{J_1^{(n)}}{F_0}\right\}
\frac{1}{G_0}
-\frac{1}{(k^2)^2}\theta(\pi^2-k^2)\delta_{n,0}
\right]
-\log \pi^2\delta_{n,0} ,\nn
\end{eqnarray}
where
\begin{eqnarray}
I_1^{(0)} &=& 4\Delta_1\Delta_3+4r^2\Delta_1^3+r^2 c\Delta_1^2
+4(c-4r^2\Delta_1-2c_\mu)\Delta_{1,0}^\mu ,
\nn\\
J_1^{(0)} &=& (\Delta_4+8r^2\Delta_1\Delta_3)(2r^2\Delta_1^2
+4\Delta_3-8\Delta_{1,0}^\mu)
+4(c_\mu +2r^2\Delta_1)(4s_\mu^2\Delta_{1,0}^\mu-\Delta_{1,1}^\mu) ,
\nn \\
I_1^{(1)} &=& 2r\left[-4\Delta_3^2+\Delta_{1,1}^\mu
+(16\Delta_3-4s_\mu^2)\Delta_{1,0}^\mu
\right] ,
\nn\\
J_1^{(1)} &=& 4r\Delta_1\left[
(\Delta_4+8r^2\Delta_1\Delta_3)(\Delta_3-4\Delta_{1,0}^\mu)
+(c_\mu +2r^2\Delta_1)(4s_\mu^2\Delta_{1,0}^\mu-\Delta_{1,1}^\mu)
\right] ,
\nn \\
I_1^{(2)} &=& \Delta_3\Delta_4-c\Delta_3^2
+\left(\frac{1}{4}c-c_\mu\right)\Delta_{1,1}^\mu
+\left[(4c_\mu-c)s_\mu^2-2\Delta_4+4\Delta_3(c-2c_\mu)\right]\Delta_{1,0}^\mu ,
\nn\\
J_1^{(2)} &=& (\Delta_4+8r^2\Delta_1\Delta_3)\left[
-2\Delta_3^2+\frac{1}{2}\Delta_{1,1}^\mu+2(4\Delta_3-s_\mu^2)\Delta_{1,0}^\mu
\right] .
\nn 
\end{eqnarray}

For quark bilinear operators,  
we parametrize $V_\Gamma$, $V_{\Gamma^\otimes}$ and
$V_{\Gamma^\prime}$ as follows:
\begin{eqnarray}
V_\Gamma &=& A^{(1)} + h_1(\Gamma) A^{(2)}+h_2(\Gamma)A^{(3)}
          +  c_{SW} h_3(\Gamma)A^{(4)} 
          + c_{SW}^2\left[h_4(\Gamma)A^{(5)}+h_1(\Gamma)A^{(6)}\right] ,
\nn\\
V_{\Gamma^\otimes} &=& \biggl(1+\frac{1}{4}h_1(\Gamma)\biggr)B^{(1)} + 
\biggl(h_2(\Gamma)-16\biggr) B^{(2)}
+  c_{SW}\left[ h_3(\Gamma)B^{(3)}
+\biggl(1+\frac{1}{4}h_1(\Gamma)\biggr)B^{(4)} \right]  
\nn\\
&+& c_{SW}^2\biggl(h_1(\Gamma)-\frac{1}{3}h_4(\Gamma)\biggr)B^{(5)} ,
\nn\\
V_{\Gamma^\prime} &=& C^{(1)}+\frac{1}{4}h_1(\Gamma)C^{(2)} + 
\frac{1}{12}\biggl(h_2(\Gamma)-16\biggr)C^{(3)}+\frac{1}{12}h_4(\Gamma)C^{(4)}
+  c_{SW}\left[\frac{1}{4}h_1(\Gamma)C^{(5)}
+\frac{1}{12}h_3(\Gamma) C^{(6)} \right]
\nn\\
&+& c_{SW}^2\left[\frac{h_1(\Gamma)}{4}C^{(7)}
+\frac{h_4(\Gamma)}{12}C^{(8)}\right] .
\nn
\end{eqnarray}
where $h_i(\Gamma) (i=1, \cdots, 4)$ are given in 
Table~\ref{tab:ds}. 
The explicit form of $A^{(n)}$ is given by
\begin{eqnarray}
A^{(n)} = 16\pi^2\int \left[
\frac{1}{F_0^2}a^{(n)}\frac{1}{G_0}
-\frac{1}{(k^2)^2}\theta(\pi^2-k^2)\delta_{n,3}
\right]
+\log \pi^2\delta_{n,3} ,\nn
\end{eqnarray}
where
\begin{eqnarray}
a^{(1)} &=& 16\left(\Delta_3+r^2\Delta_1^2\right)^2 
+\frac{16}{3}\left\{\Delta_3^2 
-\Delta_{1,1}^\mu+4\left(s_\mu^2-\Delta_3\right)\Delta_{1,0}^\mu\right\} ,
\nn\\
a^{(2)}&=&4r^2\Delta_1^2\left[\Delta_3-4\Delta_{1,0}^\mu\right] , 
\nn\\
a^{(3)}&=&\frac{1}{3}\left[-4\Delta_3^2+\Delta_{1,1}^\mu +
\left(16\Delta_3-4s_\mu^2\right)\Delta_{1,0}^\mu\right]
=\frac{1}{6r}I_1^{(1)} , 
\nn\\
a^{(4)}&=&-2r\Delta_1\cdot a^{(3)} , 
\nn\\
a^{(5)}&=&-r^2\Delta_1^2\cdot a^{(3)} , 
\nn\\
a^{(6)}&=&-3\Delta_3\cdot a^{(3)} .
\nn
\end{eqnarray}
The $B^{(n)}$ terms have the form 
\begin{eqnarray}
B^{(n)} = 16\pi^2\int 
\frac{1}{F_0}\left\{
\frac{b_0^{(n)}}{F_0}+b_1^{(n)}
\right\}\frac{1}{G_0}
\nn
\end{eqnarray}
with 
\begin{eqnarray}
b_0^{(1)}=& 32\Delta_3\left[
-r\Delta_3\Delta_1 + 4r\Delta_1 \Delta_{1,0}^\mu
\right],\quad
&b_1^{(1)}= 2r\Delta_1\left(4\Delta_3-16\Delta_{1,0}^\mu\right) ,
\nn\\
b_0^{(2)}=& 2r\Delta_1\cdot a^{(3)},\quad
&b_1^{(2)}= 0 ,
\nn\\
b_0^{(3)}=& 2(\Delta_3-r^2\Delta_1^2)\cdot a^{(3)},\quad
&b_1^{(3)}= 0 , 
\nn\\
b_0^{(4)}=& 0 ,\quad
&b_1^{(4)}= 6a^{(3)} , 
\nn \\
b_0^{(5)}=& -6r\Delta_1\Delta_3\cdot a^{(3)},\quad
&b_1^{(5)}= 0 ,
\nn 
\end{eqnarray}
and the terms $C^{(n)}$ are written as
\begin{eqnarray}
C^{(n)} &=& -16\pi^2r^2\int 
\frac{1}{F_0}\left\{\frac{c_0^{(n)}}{F_0}+c_1^{(n)}
\right\}\frac{1}{G_0} , 
\nn
\end{eqnarray}
where
\begin{eqnarray}
c_0^{(1)} =& -16 r^2 \Delta_1^2\Delta_3^2, \quad
& c_1^{(1)} = 0 , 
\nn\\
c_0^{(2)} =& 16r^2\Delta_1^2\left[
r^2 \Delta_3\Delta_1^2-4r^2\Delta_1^2\Delta_{1,0}^\mu\right],\quad
& c_1^{(2)} = 0 , 
\nn\\
c_0^{(3)} =& 4r^2\Delta_1^2
\left[-4\Delta_3^2+\Delta_{1,1}^\mu-4s_\mu^2\Delta_{1,0}^\mu
\right],\quad
& c_1^{(3)} = 0 , 
\nn\\
c_0^{(4)} =& -64r^2\Delta_1^2\Delta_3\Delta_{1,0}^\mu
,\quad
& c_1^{(4)} = 0 , 
\nn\\
c_0^{(5)} =& 0,\quad
& c_1^{(5)} = 6r\Delta_1 \cdot a^{(3)}
\nn\\
c_0^{(6)} =& 24r\Delta_1\Delta_3\cdot  a^{(3)},\quad
& c_1^{(6)} = 0 , 
\nn\\
c_0^{(7)} =& -12r^2\Delta_1^2\Delta_3\cdot a^{(3)},\quad
& c_1^{(7)} = 0 , 
\nn\\
c_0^{(8)} =& -12\Delta_3^2\cdot a^{(3)} ,\quad
& c_1^{(8)} = 0 .
\nn
\end{eqnarray}

\section*{Appendix B}

In this appendix we collect results for the one-loop relation between the 
$\overline{MS}$ and the bare coupling constant.  This relation has a 
general form 
\begin{eqnarray}
\frac{1}{g^2_{\overline{MS}}(\mu)} = \frac{1}{g^2}+(d_g+\frac{22}{16\pi^2}
\log(\mu a) ) + N_f ( d_f -\frac{4}{48\pi^2}\log(\mu a) ), 
\nn 
\end{eqnarray}
where $d_g$ only depends on the gauge action
\cite{Hasenfratz81,Dashen81,Kawai81,Weisz81,Ukawa84,Iwasaki84} 
(see Table~\ref{tab:coupling}), 
and $d_f$ only depends on the fermion action :
$d_f = 0.0066949 $ for massless Wilson quark action\cite{Kawai81} and
$d_f = 0.0314917 $ for massless clover quark action with 
$c_{SW}=1$\cite{Sint96}.
Using the average value of plaquette at one-loop, 
\begin{eqnarray}
P = 1 - c_p g^2 , 
\nn
\end{eqnarray}
we may rewrite the relation as\cite{ElKhadra92}
\begin{eqnarray}
\frac{1}{g^2_{\overline{MS}}(\mu)} = \frac{P}{g^2}+(d_g+c_p +\frac{22}{16\pi^2}
\log(\mu a) ) + N_f ( d_f -\frac{4}{48\pi^2}\log(\mu a) )  . 
\nn
\end{eqnarray}
Numerical values of $c_p$ are also collected in Table~\ref{tab:coupling}.

%%%%%%%%%%%%%%%%%%%%%%%%%%%%%%%%%%%%%%%%%%%%%%%%%%%%%%%%%%%%%%%%

\begin{table}[bht]
\caption{Finite constants for quark self-energy. 
Coefficients of the term $c_{SW}^n (n=0,1,2)$ are given in the column 
marked as $(n)$. Tadpole contributions are also listed. }
\label{tab:self}
\begin{center}
\begin{tabular}{ll|llll|llll|lll|l}
\multicolumn{2}{c|}{gauge action}&
\multicolumn{4}{c|}{$\Sigma_0$} &
\multicolumn{4}{c|}{$\Sigma_1$} &
\multicolumn{3}{c|}{$\Sigma_2$} &
$\Sigma_\alpha$ \\
$c_1$ & $c_{23}$ & 
(0) & tad& (1) & (2)  &
(0) & tad& (1) & (2)  &
(0) & (1) & (2)  &
\\
\hline
0      & 0     &-51.435 &-48.932 &13.733 &5.715 
&13.352&12.233 &-2.249 &-1.397 &
-2.100 &-9.987 &-0.017 & -4.793\\
-1/12  & 0     &-40.444 &-40.518& 11.948 &4.663 
& 9.731&10.130 &-2.015 &-1.242 &
-2.376 &-8.851 & 0.125 & \\
-0.331 & 0     &-26.073 &-29.928& 9.015 &3.106 
& 4.825 &7.482 &-1.601 &-0.973 &
-2.533 &-6.902 & 0.293 &  \\
-0.27  & -0.04 &-27.214 &-30.754& 9.283 &3.219
 & 5.208 &7.689 &-1.644 &-1.005 &
-2.552 &-7.098 & 0.287 & \\
-0.252 & -0.17 &-24.688 &-28.937& 8.698 &2.884 
& 4.294 &7.234 &-1.566 &-0.963 &
-2.572 &-6.731 & 0.324 & \\
\hline
\multicolumn{2}{c|}{$\overline{MS}$}&
0 & - & - & - & 1/2 & - & - & - & -2 & - & - &-1\\
\end{tabular}
\end{center}
\end{table}

\begin{table}[bht]
\caption{Finite part $z_m$ of renormalization factor for quark mass. 
Coefficients of the term $c_{SW}^n (n=0,1,2)$ are given in the column 
marked as $(n)$. }
\label{tab:qmass}
\begin{center}
\begin{tabular}{ll|lll}
\multicolumn{2}{c|}{gauge action}&
\multicolumn{3}{c}{$z_m$} \\
$c_1$ & $c_{23}$ & 
(0) & (1) & (2)  \\
\hline
0      & 0     &12.953 & 7.738 & -1.380 \\
-1/12  & 0     & 9.607 & 6.835 & -1.367 \\
-0.331 & 0     & 4.858 & 5.301 & -1.267 \\
-0.27  & -0.04 & 5.260 & 5.454 & -1.292 \\
-0.252 & -0.17 & 4.366 & 5.166 & -1.287 \\
\end{tabular}
\end{center}
\end{table}

\begin{table}[bht]
\caption{Finite part $V_\Gamma$ for local operators. 
Coefficients of the term $c_{SW}^n (n=0,1,2)$ are given in the column 
marked as $(n)$. Terms proportional to $c_{SW}^1$ are zero for 
pseudoscalar $P$.     
Values for $\overline{MS}$ scheme are for anti-commuting 
and 'tHooft-Veltman definition of $\gamma_5$ (latter in parentheses). }
\label{tab:local}
\begin{center}
\begin{tabular}{ll|lll|lll}
\multicolumn{2}{c|}{gauge action}&
\multicolumn{3}{c|}{$V$} &
\multicolumn{3}{c}{$A$} \\
$c_1$ & $c_{23}$ & 
(0) & (1) & (2)  &
(0) & (1) & (2)  \\
\hline
0      & 0     & 7.265&-2.497& 0.854& 2.444& 2.497&-0.854\\
-1/12  & 0     & 6.872&-2.213& 0.778& 2.808& 2.213&-0.778\\
-0.331 & 0     & 6.275&-1.725& 0.637& 3.367& 1.725&-0.637\\
-0.27  & -0.04 &6.332&-1.775&0.652&3.315&1.775&-0.652\\
-0.252 & -0.17 &6.231&-1.683&0.625&3.414&1.683&-0.625\\
\hline
\multicolumn{2}{c|}{$\overline{MS}$}&
-1/2 & - & - & -1/2 & - & - \\
&&
(-1/2) & - & - & (-9/2) & - & - \\
\end{tabular}

\begin{tabular}{ll|lll|ll|lll}
\multicolumn{2}{c|}{gauge action}&
\multicolumn{3}{c|}{$S$} &
\multicolumn{2}{c|}{$P$} & 
\multicolumn{3}{c}{$T$} \\
$c_1$ & $c_{23}$ & 
(0) & (1) & (2)  &
(0) & (2)  &
(0) & (1) & (2)  \\
\hline
0      & 0     &
 2.100& 9.987& 0.017& 11.743&  3.433& 4.166&-1.665&-0.575\\
-1/12  & 0     &
2.376&8.851&-0.125&10.502&2.987&4.307&-1.475&-0.477\\
-0.331 & 0     &
 2.533& 6.902&-0.293& 8.348&  2.254& 4.615&-1.150&-0.327 \\
-0.27  & -0.04 &
2.552&7.098&-0.287&8.584&2.321&4.575&-1.183&-0.339 \\
-0.252 & -0.17 &
2.572&6.731&-0.324&8.208&2.175&4.633&-1.122&-0.309\\
\hline
\multicolumn{2}{c|}{$\overline{MS}$}&
2 & - & - & 2 & - & 0 & - & -\\
&&
(2) & - & - & (10) & - & (0) & - & -\\
\end{tabular}
\end{center}
\end{table}

\begin{table}[bht]
\caption{Finite part $V_{\Gamma^\otimes}$ for improved operators.
Coefficients of the term $c_{SW}^n (n=0,1,2)$ are given in the column 
marked as $(n)$. Values for pseudoscalar $P$ are zero.  }
\label{tab:impr}
\begin{center}
\begin{tabular}{ll|lll|lll|lll|lll}
\multicolumn{2}{c|}{gauge action}&
\multicolumn{3}{c|}{$V$} &
\multicolumn{3}{c|}{$A$} &
\multicolumn{3}{c|}{$S$} &
\multicolumn{3}{c}{$T$} \\
$c_1$ & $c_{23}$ & 
(0) & (1) & (2)  &
(0) & (1) & (2)  &
(0) & (1) & (2)  &
(0) & (1) & (2)  \\
\hline
0      & 0     & 
-9.786 &   3.416 &    0.885 & -19.372 &  10.317 &   -0.885 &
-19.172 &  13.801 &  -3.538 & 
-16.244 &   6.855 &    0.590 \\
-1/12   & 0     &
 -8.234 &   3.112 &    0.756 & -15.852 &   8.836 &   -0.756 &
-15.235 &  11.448 &  -3.025 & 
-13.518 &   6.058 &    0.504 \\
-0.331 & 0     &
 -5.881 &   2.547 &    0.546 & -10.741 &   6.468 &   -0.546 &
 -9.720 &   7.842 &  -2.183 & 
 -9.461 &   4.703 &    0.364 \\
-0.27  & -0.04 &
 -6.085 &   2.608 &    0.564 & -11.156 &   6.675 &   -0.564 &
-10.143 &   8.135 &  -2.257 & 
 -9.804 &   4.833 &    0.376 \\
-0.252 & -0.17 &
 -5.640 &   2.498 &    0.521 & -10.189 &   6.200 &   -0.521 &
 -9.098 &   7.403 &  -2.084 & 
 -9.036 &   4.565 &    0.347 \\
\end{tabular}
\end{center}
\end{table}

\begin{table}[bht]
\caption{Finite part $V_{\Gamma^\prime}$ for improved operators. 
Coefficients of the term $c_{SW}^n (n=0,1,2)$ are given in the column 
marked as $(n)$. }
\label{tab:impr_a2}
\begin{center}
\begin{tabular}{ll|lll|lll}
\multicolumn{2}{c|}{gauge action}&
\multicolumn{3}{c|}{$V$} &
\multicolumn{3}{c}{$A$} \\
$c_1$ & $c_{23}$ & 
(0) & (1) & (2)  &
(0) & (1) & (2)  \\
\hline
0      & 0     & -3.525 &  1.973 &  -0.412 &  3.296 & -1.973 &   0.412 \\
-1/12  & 0     &-2.717 &  1.575 &  -0.338 &  2.549 & -1.575 &   0.338 \\
-0.331 & 0     &-1.646 &  1.007 &  -0.225 &  1.552 & -1.007 &   0.225 \\
-0.27  & -0.04 &-1.715 &  1.045 &  -0.233 &  1.613 & -1.045 &   0.233 \\
-0.252 & -0.17 &-1.496 &   0.921 &  -0.209 &  1.402 &  -0.921 &   0.209 \\
\end{tabular}

\begin{tabular}{ll|lll|lll|lll}
\multicolumn{2}{c|}{gauge action}&
\multicolumn{3}{c|}{$S$} &
\multicolumn{3}{c|}{$P$} &
\multicolumn{3}{c}{$T$} \\
$c_1$ & $c_{23}$ & 
(0) & (1) & (2)  &
(0) & (1) & (2)  &
(0) & (1) & (2)  \\
\hline
0      & 0     &
 4.981 & -2.177 &   0.288 & -8.660 &  5.715 & -1.361 &   0.461 &  -0.590 
&  0.179 \\
-1/12  & 0     &
3.751 & -1.638 &   0.205 & -6.780 &  4.663 & -1.145 &   0.393 &  -0.504
&  0.157 \\
-0.331 & 0     &
 2.171 &  -0.923 &   0.097 & -4.225 &  3.106 &  -0.804 &   0.280 &  -0.364 
&  0.118 \\
-0.27  & -0.04 &
 2.260 &  -0.962 &   0.102 & -4.396 &  3.219 &  -0.831 &   0.288 &  -0.376
&  0.121 \\
-0.252 & -0.17 &
 1.925 &  -0.801 &   0.078 & -3.870 &  2.884 &  -0.758 &   0.261 &  -0.347 
&  0.113 \\
\end{tabular}
\end{center}
\end{table}

\begin{table}[bht]
\caption{Finite part $z_\Gamma$ of renormalization factor for local 
bilinear quark operators 
($z=0$ in (\protect\ref{eq:operator}) and (\protect\ref{eq:renop})).
Coefficients of the term $c_{SW}^n (n=0,1,2)$ are given in the column 
marked as $(n)$. }
\label{tab:zfact}
\begin{center}
\begin{tabular}{ll|lll|lll}
\multicolumn{2}{c|}{gauge action}&
\multicolumn{3}{c|}{$V$} &
\multicolumn{3}{c}{$A$} \\
$c_1$ & $c_{23}$ & 
(0) & (1) & (2)  &
(0) & (1) & (2)  \\
\hline
0      & 0     & 30.817 & -8.988 &  -5.172 & 
35.638 & -13.981 & -3.464 \\
-1/12  & 0     & 23.841 & -7.720 & -4.199 & 
27.904 & -12.146 & -2.642 \\
-0.331 & 0     & 14.974 & -5.689 & -2.770 &
17.881 & -9.140 & -1.496 \\
-0.27  & -0.04 & 15.674 & -5.865 & -2.866 &
18.691 & -9.414 & -1.562 \\
-0.252 & -0.17 & 14.163 & -5.450 & -2.546 &
16.981 & -8.815 & -1.297 \\
\end{tabular}

\begin{tabular}{ll|lll|lll|lll}
\multicolumn{2}{c|}{gauge action}&
\multicolumn{3}{c|}{$S$} &
\multicolumn{3}{c|}{$P$} & 
\multicolumn{3}{c}{$T$} \\
$c_1$ & $c_{23}$ & 
(0) & (1) & (2)  &
(0) & (1) & (2)  &
(0) & (1) & (2)  \\
\hline
0      & 0     & 38.482 & -21.471 & -4.335 &
28.839 & -11.484 & -7.751 & 34.417 & -9.820 & -3.743 \\
-1/12  & 0     & 30.837 & -18.784 & -3.296 &
22.710 & -9.933 & -6.408 & 26.905 & -8.458 & -2.944 \\
-0.331 & 0     & 21.215 & -14.316 & -1.840 &
15.400 & -7.414 & -4.387 & 17.134 & -6.264 & -1.806 \\
-0.27  & -0.04 & 21.954 & -14.737 & -1.927 &
15.922 & -7.639 & -4.535 & 17.931 & -6.456 & -1.875 \\
-0.252 & -0.17 & 20.322 & -13.864 & -1.597 &
14.687 & -7.133 & -4.096 & 16.261 & -6.011 & -1.613 \\
\end{tabular}
\end{center}
\end{table}

\begin{table}[bht]
\caption{Factors for $\gamma$ matrix contractions. 
Values in parentheses are for 'tHooft-Veltman definition of $\gamma_5$; 
others are for anti-commuting $\gamma_5$. }
\label{tab:ds}
\begin{center}
\begin{tabular}{l|lllll|llll}
$\Gamma$ & $h_1(\Gamma)$ & $h_2(\Gamma)$ & $h_3(\Gamma)$ &
$h_4(\Gamma)$ & $j(\Gamma)$ & $3h_2/8+j$ & $1+h_1/4$ &$h_2-16$ & $h_1-h_4/3$\\
\hline
$\gamma_\mu\gamma_5$ & 2 & 4 & -6 &   0 & -2(-6)& -1/2(-9/2)&3/2&-12& 2\\
$\gamma_\mu$         &-2 & 4 &  6 &   0 & -2& -1/2&1/2&-12&-2\\
$\gamma_5$           &-4 &16 &  0 & -12 & -4(4) &  2(10)  &0  &0  &0\\
1                    & 4 &16 &-24 & -12 & -4&  2  &2  &0  &8\\
$\sigma_{\mu\nu}$    & 0 & 0 &  4 &   4 &  0 &  0  &1  &-16&-4/3\\
\end{tabular}
\end{center}
\end{table}

\begin{table}
\caption{One-loop corrections for coupling constant $d_g, d_f$ and plaquette
$c_p$. }
\label{tab:coupling}
\begin{center}
\begin{tabular}{ll|ll}
\multicolumn{2}{c|}{gauge action}&
$d_g$ & $c_p$ \\
$c_1$ & $c_{23}$ & & \\
\hline
0      & 0    &-0.4682 & 1/3    \\
-1/12  & 0    &-0.2361 & 0.2442 \\
-0.331 & 0    & 0.1000 & 0.1402 \\
-0.27  &-0.04 &        & 0.1472 \\
-0.252 &-0.17 & 0.1196 & 0.1286 \\
\end{tabular}
\end{center}
\end{table}

%%%%%%%%%%%%%%%%%%%%%%%%%%%%%%%%%%%%%%%%%%%%%%%%%%%%%%%%%%%%%%%%%%%%%%%%%%

\end{document}